\documentclass[aps,pra,twocolumn,showpacs,amsmath,amssymb,floatfix,superscriptaddress]{revtex4-1}
\usepackage{graphicx}
\usepackage[bookmarks,bookmarksopen,bookmarksnumbered,colorlinks,linkcolor=red,linktocpage,citecolor=blue,urlcolor=cyan,pdfpagemode=UseOutline]{hyperref}

\def\ben{\begin{equation}}
\def\een{\end{equation}}
\def\sss{\scriptscriptstyle\rm}
\def\xc{_{\sss XC}}
\def\br{{\bf r}}
\newcommand{\parref}[1]{(\ref{#1})}
\newcommand{\matelem}[3]{\left\langle #1 \left| #2 \right| #3 \right\rangle}
\providecommand{\abs}[1]{\left|#1\right|}
\providecommand{\bra}[1]{\left< #1 \right|}
\providecommand{\ket}[1]{\left| #1 \right>}

\begin{document}

\title{Full self-consistency in Fermi-orbital self-interaction correction}
\author{Zeng-hui Yang}
\affiliation{Department of Physics, Temple University, Philadelphia, PA 19122, USA}
\altaffiliation{Current address: Microsystem and Terahertz Research Center, China Academy of Engineering Physics, Chengdu, China 610200}
\author{Mark R. Pederson}
\affiliation{Department of Chemistry, John Hopkins University, Baltimore, MD 21218, USA}
\author{John P. Perdew}
\affiliation{Department of Physics, Temple University, Philadelphia, PA 19122, USA}
\date{\today}

\begin{abstract}
The Perdew-Zunger self-interaction correction cures many common problems associated with semilocal density functionals, but suffers from a size-extensivity problem when Kohn-Sham orbitals are used in the correction. Fermi-L\"{o}wdin-orbital self-interaction correction (FLOSIC) solves the size-extensivity problem, allowing its use in periodic systems and resulting in better accuracy in finite systems. Although the previously published FLOSIC algorithm [J. Chem. Phys. 140, 121103 (2014)] appears to work well in many cases, it is not fully self-consistent. This would be particularly problematic for systems where the occupied manifold is strongly changed by the correction. In this paper we demonstrate a new algorithm for FLOSIC to achieve full self-consistency with only marginal increase of computational cost. The resulting total energies are found to be lower than previously reported non-self-consistent results.
\end{abstract}

\maketitle

\section{Introduction}
The Kohn-Sham\cite{KS65} (KS) density-functional theory\cite{KS65,HK64,FNM03} (DFT) is a widely used computational method in physics and chemistry. To perform a DFT calculation, one needs to approximate the so-called exchange-correlation (xc) energy, a functional of the electronic density. Most of the available approximations are semilocal approximations, where the xc energy density at a certain point only depends on the density and derivatives of the density at that point. These approximations have been extensively used thanks to their computational efficiency and they generally work well, but they all contain the self-interaction error (SIE), which means that the sum of the Hartree interaction energy and the approximated xc energy does not properly vanish for all one-electron systems. SIE causes a wide range of problems, such as the xc potential decaying too fast asymptotically, the orbital energies of occupied orbitals lying too high in a non-systematic way, and the total energy varying in a strongly nonlinear way between adjacent integer electron numbers. These problems show up in DFT calculations as missing Rydberg states, mis-ordering of states for some systems, incorrect description of stretched bonds, problems with band gaps, unstable anions, and similar uncertainties in modeling processes that depend upon such phenomena.

The self-interaction correction\cite{PZ81} (SIC), proposed by Perdew and Zunger (PZ), cures these problems by introducing orbital-dependent corrections to the xc energy functional:
\ben
E\xc^\text{SIC}[n_\uparrow,n_\downarrow]=E\xc[n_\uparrow,n_\downarrow]-\sum_\sigma\sum_{i}^{N_\sigma}\left\{U[n_{i\sigma}]+E\xc[n_{i\sigma},0]\right\},
\label{eqn:PZSIC}
\een
where $\sigma$ is the spin index, $N_\sigma$ is the number of occupied orbitals of spin $\sigma$, and $n_{i\sigma}=|\psi_{i\sigma}|^2$ is the orbital density of the KS orbital $\psi_{i\sigma}$. Eq. \parref{eqn:PZSIC} can be directly applied to any xc energy functional, and in principle an approximated density functional could be developed within a systematic self-interaction corrected formalism. The correction vanishes when $E\xc$ is the exact xc functional. Eq. \parref{eqn:PZSIC} corrects the SIE, but it can also deviate from the KS scheme of DFT, which only allows a local multiplicative potential for the system. While through the optimized effective potential (OEP)\cite{SH53,TS76} technique the SIC calculation can be done within the KS scheme\cite{KKM08}, the cost of doing so can outweigh the benefit. Also, a recent study\cite{YPSP16} supports the conclusion that properties such as the band gap are inherently inaccurate in the KS scheme, so there is less reason to use the OEP procedure in practical calculations. The commonly used SIC functional is thus defined in the generalized KS scheme\cite{SGVM96}. The computational cost of PZSIC is higher than that of regular DFT, since the problem has yet to be reduced to a simple eigenvalue problem.

PZSIC works well for small molecules, but there is a flaw that the correction for extended systems can vanish in the bulk limit\cite{PZ81} if the SIC energy is evaluated naively with KS orbitals. The reason is easy to see. For periodic systems, the KS orbitals all contain a normalization constant of $1/\sqrt{V_\text{crystal}}$. When using KS orbitals in Eq. \parref{eqn:PZSIC}, the magnitude of the correction per electron decreases as the size of the crystal increases, and eventually vanishes in the limit of an infinitely large crystal. The problem also shows up in stretched bonds and other situations\cite{KKJ11}. This is the so-called size-extensivity problem, that the energy of separated fragments does not equal the sum of energies of the fragments as it should. For this reason, Perdew and Zunger suggested that the SIC functional could be constructed in terms of a set of possibly-localized orbitals that minimize the SIC energy. However, this only solves the problem when each SIC term in the energy is negative\cite{PZ81} after localization. This is a problem with the original PZSIC formulation.

The total energy of a regular DFT calculation is invariant with respect to unitary rotations of KS orbitals, but this is no longer the case for SIC calculations, since the SIC term depends explicitly on orbitals. By applying a unitary rotation that transforms occupied KS orbitals to localized orbitals, one can often lower the energy and achieve a finite correction in periodic systems. Localization schemes have been explored thoroughly by the Lin group\cite{HHL83,PHL84,PHL85,PL88}, where the localized orbitals that minimize the energy are obtained by solving the following localization equation\cite{PHL84,PHL85}
\ben
\matelem{\phi_{j\sigma}}{V^\text{SIC}_{j\sigma}-V^\text{SIC}_{i\sigma}}{\phi_{i\sigma}}=0
\label{eqn:localizationEquation}
\een
together with the KS-SIC equation of localized orbitals:
\ben
\left(\hat{H}_\sigma^\text{KS}+V_{i\sigma}^\text{SIC}\right)\phi_{i\sigma}=\sum_j^{N_\sigma}\lambda_{ji\sigma}\phi_{j\sigma},
\label{eqn:localizedOrbitalSIC}
\een
where $\phi$ denotes a localized orbital, $\sigma$ is the spin index, $V^\text{SIC}_{i\sigma}(\br)=-\delta\left\{U[n]+E\xc[n,0]\right\}/\delta n(\br)|_{n(\br)=\abs{\phi_{i\sigma}(\br)}^2}$ for occupied orbitals, and $V^\text{SIC}_{i\sigma}(\br)=0$ for virtual orbitals. We denote the canonical orbitals as $\psi$, which diagonalize the Lagrange multiplier matrix $\lambda_\sigma$ (the diagonal elements are orbital energies in SIC) and are the analog of KS orbitals in SIC. The sets of $\psi$ and $\phi$ are related to each other by a unitary transformation. Eq. \parref{eqn:localizedOrbitalSIC} can also be formulated in terms of canonical orbitals, but it is computationally preferable to solve for the localized orbitals directly instead of calculating the canonical orbitals first. The set of localized orbitals that satisfy Eq. \parref{eqn:localizationEquation} and \parref{eqn:localizedOrbitalSIC} is not unique, and in practice one has to be careful in choosing the initial guess when solving these equations to avoid converging to a local minimum. The localization equation method works generally well, but the computation cost is prohibitively high, as it scales as $O(N^6)$in the small N limit\cite{PRP14}. As a comparison, the coupled cluster singles doubles (CCSD)\cite{PB82} method has a similar scaling, but it is typically more accurate than DFT methods. Therefore the localization equation method has only been applied to small systems, and it has little advantage compared to methods of similar scaling.

The Fermi-L\"{o}wdin-orbital (FLO) SIC \cite{PRP14} is a recently proposed localization scheme, which circumvents the need for solving Eq. \parref{eqn:localizationEquation}. Instead of searching over all possible unitary transformations to minimize the energy, FLOSIC restricts the transformation to a very specific form, where the localized orbitals are derived directly from the so-called Fermi orbitals:
\ben
\phi_{i\sigma}^\text{FO}(\br)=\frac{n_\sigma(\mathbf{a}_{i\sigma},\br)}{\sqrt{n_\sigma(\mathbf{a}_{i\sigma})}},
\label{eqn:FO}
\een
where $n_\sigma(\mathbf{a}_{i\sigma},\br)=\sum_j^{N_\sigma}\psi_{j\sigma}^*(\mathbf{a}_{i\sigma})\psi_{j\sigma}(\br)$ is the single-particle density matrix of the KS system, and $\mathbf{a}_{i\sigma}$ is the Fermi orbital descriptor (FOD). The localized orbitals $\{\phi_{i\sigma}\}$ are the L\"{o}wdin-orthogonalized\cite{PHL85,L62} Fermi orbitals $\{\phi_{i\sigma}^\text{FO}\}$. The unitary transformation between canonical orbitals and localized orbitals is determined with only $N_\sigma$ parameters ($\mathbf{a}_{i\sigma}$) instead of $N_\sigma^2$ parameters (the unitary transformation matrix). The FODs can be seen as quasi-classical electron positions, and they usually can be reliably guessed based on physical and symmetry arguments. Therefore, even though the set of FODs that minimizes the energy can also be non-unique, the situation is much better than in the localization equation method.

Motivated by earlier work~\cite{HHL83}, the FLOSIC calculation in its original paper Ref. \cite{PRP14} is realized by adding an extra matrix to the Kohn-Sham Hamiltonian $\hat{H}_\sigma^\text{KS}$ and diagonalizing the resulting matrix $\hat{H}_\sigma$:
\ben
\hat{H}_\sigma=\hat{H}_\sigma^\text{KS}+\sum_{ij}^{N_\sigma}\frac{V_{ij}^{(i\sigma)}+V_{ij}^{(j\sigma)}}{2}\ket{\phi_{i\sigma}}\bra{\phi_{j\sigma}},
\label{eqn:FOSICold}
\een
where $V_{ij}^{(i\sigma)}=\matelem{\phi_{i\sigma}}{V_{i\sigma}^\text{SIC}}{\phi_{j\sigma}}$. This physically motivated Hermitian hamiltonian contains most of the effects due to the self-interaction-corrected functional and has been shown to yield canonical-orbital eigenvalues that are not too different from the exact canonical-orbital eigenvalues used in earlier versions of SIC, but it is an approximation~\cite{PHL85,PL88}. This algorithm has the advantage of its simplicity for implementation, since the problem remains an eigenvalue problem. However, transforming the orbital-dependent SIC to a regular eigenvalue problem introduces non-self-consistency, which is evident from Eq. \parref{eqn:FOSICold}: the canonical orbitals obtained from solving Eq. \parref{eqn:FOSICold} are restricted inside the occupied Hilbert space of the previous self-consistent-field (SCF) iteration. Even though the orbitals can still drift away from the initially occupied space since $\hat{H}_\sigma^\text{KS}$ changes between SCF iterations due to density change, the algorithm is not fully self-consistent (SC).

While non-SC FLOSIC energies, including cohesive energies, are often quite accurate if wave functions from a nearly identical functional are used, SC FLOSIC is needed to make sure that the results are reliable in difficult cases such as molecular magnets\cite{PKP06}, and to have a Hellmann-Feynman theorem which allows for the efficient optimization of molecular and unit-cell geometries.

The non-interacting spin-density matrix $n_\sigma(\br',\br)$ is the matrix in the position representation of the projection operator onto the space or manifold of the occupied orbitals of spin $\sigma$. Thus it is invariant under unitary transformation of those orbitals. We can calculate the total energy from $n_\sigma(\br',\br)$, and we can also find the electron spin-density $n_\sigma(\br)=n_\sigma(\br,\br)$ from its diagonal. We will exploit these facts here to find the optimized FLOs, for a given set of FODs directly. Before the first iteration, we will have approximate canonical occupied orbitals, which generate a set of initial FLOs, and canonical unoccupied orbitals. Then we will mix the occupied and unoccupied orbitals to lower the SIC energy, with a reconstruction of localized FLOs after each iteration. We will not need to generate canonical occupied orbitals after the iteration begins, but they can be obtained easily if needed.

\section{Algorithm}
\begin{figure}
\includegraphics[width=0.85\columnwidth]{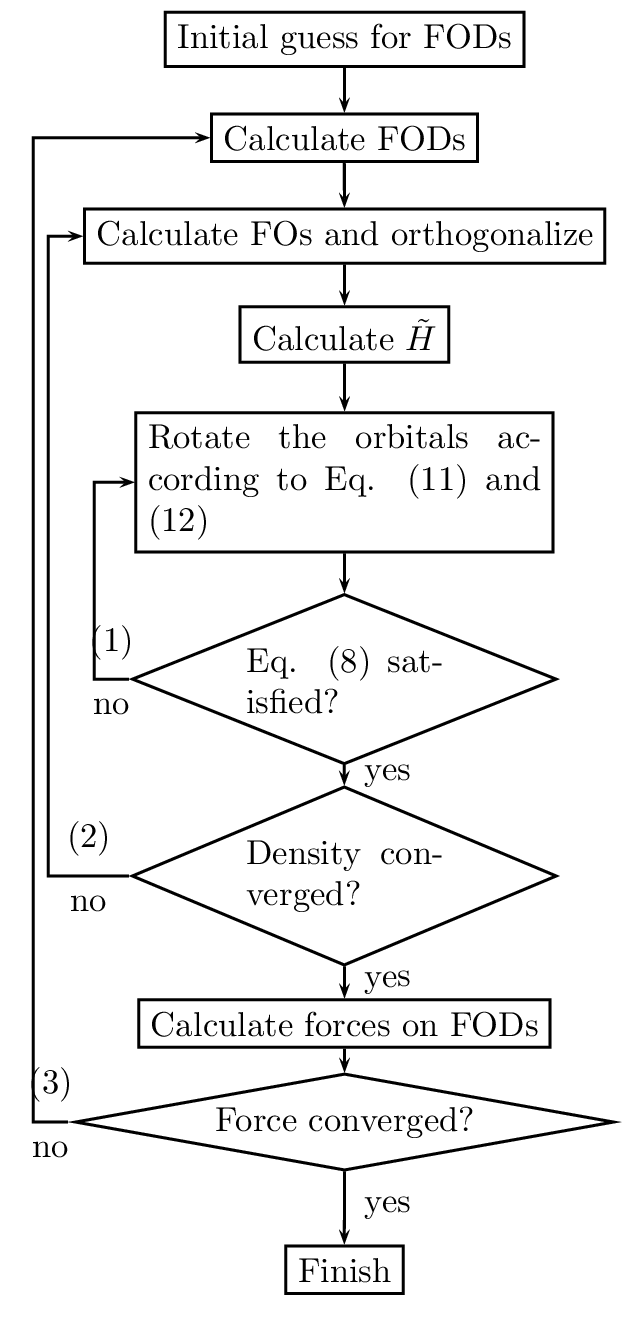}
\caption{Flow chart of the SC-FLOSIC algorithm presented in this work. Loop $(1)$ solves for the localized orbitals. Loop $(2)$ is the regular SCF loop. Loop $(3)$ minimizes the total energy with respect to the FODs.}
\label{fig:flowchart}
\end{figure}

In this work we set out to find the SC solution of Eq. \parref{eqn:localizedOrbitalSIC}, where the localized orbitals are constrained to be symmetrically orthogonalized Fermi orbitals which are referred to as Fermi-L\"{o}wdin Orbitals (FLO). Both the localized orbitals and the Lagrange multiplier matrices need to be determined. Since the Lagrange multiplier matrices are in principle different for hamiltonians defined by different orbitals, solving Eq. \parref{eqn:localizedOrbitalSIC} directly would mean determining $N_\sigma^3$ variables for a given set of FODs. Even though we only need to determine $N_\sigma$ parameters in the FLOSIC method instead of $N_\sigma^2$ parameters in the localization equation method, the computational cost is still very high. Instead of trying to solve Eq. \parref{eqn:localizedOrbitalSIC} directly, we develop a Jacobi-like method for solving Eq. \parref{eqn:localizedOrbitalSIC} iteratively. Even though we employ several approximations to reduce the computational cost, the final result obtained from the iterative procedure remains an exact solution of Eq. \parref{eqn:localizedOrbitalSIC} within the chosen basis set.

In the following, we represent all matrices in the basis set formed by the FLOs (indexed $1$ to $N_\sigma$) and the virtual KS orbitals (indexed $N_\sigma+1$ to $M$, $M$ being the size of the basis set) of the previous SCF iteration, all of which are denoted as $\phi_{i\sigma}$. Note that all $\phi_{i\sigma}$'s are orthogonal to each other. The FLOs are calculated with pre-determined FODs, which are fixed during the SCF loop (refer to Fig. \ref{fig:flowchart}). The Hamiltonian matrix of occupied orbital $i\sigma$ in this representation would be
\ben
H^{(i\sigma)}_{mn}=\matelem{\phi_{m\sigma}}{H_\sigma^\text{KS}+V^\text{SIC}_{i\sigma}}{\phi_{n\sigma}}.
\label{eqn:Hmat}
\een
To reduce the amount of calculations needed, instead of calculating Eq. \parref{eqn:Hmat} exactly, we construct a matrix for the operator $\tilde{H}_\sigma$
\ben
\tilde{H}_{mn\sigma}=\matelem{\phi_{m\sigma}}{H_\sigma^\text{KS}+V^\text{SIC}_{m\sigma}}{\phi_{n\sigma}},
\label{eqn:Htildemat}
\een
so that the $i$-th row of $\tilde{H}_\sigma$ is the same as the $i$-th row of $H^{(i\sigma)}$. According to Eq. \parref{eqn:localizedOrbitalSIC}, the upper-left $N_\sigma\times N_\sigma$ corner of $\tilde{H}_\sigma$ is the transpose of the Lagrange multiplier matrix $\lambda_\sigma$.

We have the following relations if the group of $\phi_{i\sigma}$ solves Eq. \parref{eqn:localizedOrbitalSIC}:
\begin{align}
\tilde{H}_{i\le N_\sigma,j>N_\sigma,\sigma}&=0 \label{eqn:zeros}\\
\tilde{H}_{i>N_\sigma,j>N_\sigma,\sigma}&=\epsilon_i\delta_{ij} \label{eqn:virts}.
\end{align}
Eq. \parref{eqn:zeros} holds since the Hilbert spaces spanned by the occupied and virtual orbitals are orthogonal, and Eq. \parref{eqn:virts} holds since $V^\text{SIC}_{i\sigma}=0$ for virtual orbitals.

Eq. \parref{eqn:zeros} only holds when the group of $\phi_{i\sigma}$ solves Eq. \parref{eqn:localizedOrbitalSIC}. Consider the case where trial functions $\hat{\phi}_{i\le N_\sigma,\sigma}$ are unitarily-transformed $\phi_{i\sigma}$:
\ben
\hat{\phi}_{i\sigma}=\sum_{j}^{N_\sigma}C_{ij}\phi_{j\sigma},\quad i\le N_\sigma.
\een
Unless $C_{ij}=\delta_{ij}$, applying $\hat{H}_\sigma^\text{KS}+V_{i\sigma}^\text{SIC}$ on $\hat{\phi}_{i\sigma}$ will generally yield $\sum_{j}^{N_\sigma}\hat{\lambda}_{ji\sigma}\phi_{j\sigma}+\Delta_i$, where the residue $\Delta_i$ belongs to the virtual Hilbert space. Therefore instead of trying to solve Eq. \parref{eqn:localizedOrbitalSIC} directly, we can solve Eq. \parref{eqn:localizedOrbitalSIC} by requiring that Eqs. \parref{eqn:zeros} and \parref{eqn:virts} be satisfied. This is done by finding a set of $\phi'_{i\sigma}$ that makes the upper-right $(M-N_\sigma)\times N_\sigma$ corner of $\tilde{H}$ vanish in the matrix representation where the group of $\phi'_{i\sigma}$ forms the basis set. We achieve this by pairwise mixing the FLOs with the virtual KS orbitals. For every pair of FLO $\phi_{i\le N_\sigma,\sigma}$ and virtual orbital $\phi_{j>N_\sigma,\sigma}$, we define $\phi'_{i\sigma}$ and $\phi'_{j\sigma}$ as a rotation of the original pair:
\ben
\begin{split}
\phi'_{i\sigma}&=\cos t\,\phi_{i\sigma}+\sin t\,\phi_{j\sigma},\\
\phi'_{j\sigma}&=-\sin t\,\phi_{i\sigma}+\cos t\,\phi_{j\sigma}.
\end{split}
\label{eqn:rotate}
\een
We then require that $\matelem{\phi'_{i\sigma}}{H_\sigma^\text{KS}+V^\text{SIC}_{i\sigma}}{\phi'_{j\sigma}}=0$ and obtain the equation for the rotation angle $t$:
\ben
t=\frac{1}{2}\arctan\left(\frac{2\tilde{H}_{ij\sigma}}{\tilde{H}_{ii\sigma}-\matelem{\phi_{j\sigma}}{H_\sigma^\text{KS}+V^\text{SIC}_{i\sigma}}{\phi_{j\sigma}}}\right).
\label{eqn:tangle}
\een
The angle $t$ being in the correct direction is more important than its actual value. Not using its exact value only impacts the speed of convergence. We also found that at times convergence can be hastened by stepping a shorter distance than determined from Eq. \parref{eqn:tangle}. Such rotations are done for all FO-virtual orbital pairs. After each rotation, we replace $\phi_{i\sigma}$ and $\phi_{j\sigma}$ with $\phi'_{i\sigma}$ and $\phi'_{j\sigma}$, so that the later rotations will take the earlier rotations into account. Convergence is checked after all pairs of orbitals are rotated, and it is reached when the biggest matrix element in the upper-right corner of $\tilde{H}$ is smaller than a pre-determined criterion. The rotation procedure is repeated until convergence is reached. We find that the updated occupied orbitals are still primarily localized to their original regions and the updated virtual orbitals are still primarily delocalized.

This Jacobi-like method for zeroing-out the upper-right corner of $\tilde{H}$ shows strong dependence on the initial set of $\phi_{i\sigma}$. For some starting points more than 1000 iterations are needed for convergence. To accelerate the convergence, we precondition the starting orbitals by diagonalizing the averaged matrix $(\tilde{H}+\tilde{H}^T)/2$, which yields a unitary transformation between approximate canonical orbitals of the current iteration and the occupied FLO and unoccupied delocalized orbitals (UDO) of the previous iteration. We then transform back to an FLO/UDO representation before performing the Jacobi-like iterative update. This method for preconditioning always converges within 20 iterations for all systems checked. This can be seen as an equivalent to applying a pre-conditioner to $\tilde{H}$. The upper-left corner of the resulting $\tilde{H}$ matrix(the new Lagrange multiplier matrix $\lambda_\sigma$) is also found to be almost diagonal, meaning that the group of $\phi$ after rotation is very close to canonical orbitals.  As discussed in greater detail below, this preconditioning approach can be, and for the results here has been, used to converge to approximate canonical orbitals or exact FLO depending on whether a back transformation scheme is used after the exact diagonalization. Working within the FLO representation will most likely be more efficient with respect to the use of order-N methods and has definite advantages for applications of SIC to meta-stable donor-acceptor complexes as demonstrated for LiF in the applications section.

Though the SIC equation only applies to occupied orbitals, it is possible to introduce orbital-dependent potentials for virtual orbitals\cite{HHL83b,HG69,SC84,BP09,JL88,PK88}. This would not change the complexity of the method, but it can affect the convergence speed. Since the relative mixing of the virtual orbitals with the occupied orbitals changes as the process iterates to self-consistency, the energy gap between the occupied and unoccupied orbitals affects the rate of self-consistency. In an early paper, Harrison, Heaton and Lin \cite{HHL83b} suggested that the unoccupied orbitals should move in a potential that resembles the SIC potential of the lowest unoccupied localized orbital since that potential is similar to the potential that would arise if the particle-hole interaction is included. This method is similar to the improved-virtual-orbital (IVO) method of Schneider and Collins \cite{SC84} and a recent rendition by Pederson and Baruah \cite{BP09}. Approaches such as these lead to more meaningful estimates of localized excitation energies (such as excitions) and allow for a pre-screening of electronically interesting geometries. They will likely decrease the number of self-consistent iterations when the gap is small but may increase the number of self-consistent iterations when the gap is big. Early efforts to use IVO-like approaches to simultaneously determine the ground-state and excited state energies in defects include work by Jackson and Lin\cite{JL88} and Pederson and Klein \cite{PK88}.

The method can be modified to produce a group of $\phi_{i\sigma}$ that is close to FLOs instead of canonical orbitals, which can be useful for ensuring consistent convergence to the same ground state in stretched bonds (Fig. \ref{fig:LiFionic} is an example) and other difficult cases. In the previously mentioned trick, the starting orbitals are the eigenvectors of $(\tilde{H}+\tilde{H}^T)/2$, and we denote them as $\tilde{\psi}_{i\sigma}$. We construct auxiliary orbitals $\tilde{\phi}_{i\sigma}$ by projection into the occupied and virtual spaces of the previous SCF iteration:
\ben
\tilde{\phi}_{i\sigma}=\left\{\begin{array}{ll}\sum_{j=1}^{N_\sigma}\left<\phi_{j\sigma}\right.\left|\tilde{\psi}_{i\sigma}\right>\ket{\phi_{j\sigma}} & i\le N_\sigma,\\
\sum_{j=N_\sigma+1}^{M}\left<\phi_{j\sigma}\right.\left|\tilde{\psi}_{i\sigma}\right>\ket{\phi_{j\sigma}}& i>N_\sigma.\end{array}\right.
\een
The new starting orbitals for the Jacobi-like method are obtained by applying the L\"{o}wdin symmetric orthogonalization\cite{PHL85,L62} separately to $\{\tilde{\phi}_{i\le N_\sigma,\sigma}\}$ and $\{\tilde{\phi}_{i>N_\sigma,\sigma}\}$. By truncating the part of $\tilde{\psi}_{i\sigma}$ that does not lie within the occupied space of the previous iteration when $i\le N_\sigma$, we make the starting orbitals as close to the FLOs of the previous iteration as possible, and we find the results after the rotations are close to the exact FLOs. With this choice of starting orbitals, the diagonal elements of the Lagrange multiplier matrix are not approximations to orbital energies anymore, as the Lagrange multiplier matrix is no longer near-diagonal. However, the total energy does not depend on the choice of the starting orbitals of the Jacobi-like method.

The algorithm described above only comprises a complete calculation when optimal FODs are known. The optimal FODs can be guessed in very simple systems based on symmetry arguments (such as in $\mathrm{H}_2$), but for most of the systems they have to be optimized to ensure that the total energy is at the lowest point. The initial FODs need to be guessed before the first SCF step. Since it is found that FODs often represent chemical bonds and lone pairs\cite{PRP14,HLJP15}, making the initial guess is usually simple in first- and second-row elements. We calculate the forces on the FODs after the SCF loop for electronic structures as described in Ref. \cite{P15}. These forces are then used as input of the conjugate-gradient method to generate new FODs, which are used in the next SCF step. This optimization of the FODs is repeated until the forces converge.

Symmetry can be used in the optimization of the FODs to simplify the process and to improve the stability of the optimization.
Although this is not a general fact, the optimal FODs for all the systems we tested transform according to a sub-group of the molecular point group. One can easily guess the point group of the FODs for simple systems such as the ones in Table \ref{table:res}, but currently there does not exist a completely systematic way of determining the symmetry of FODs for more complex systems before the calculation.

In regular KS calculations, the KS potential is usually mixed with that of the previous iteration for numerical stability. The non-SC FLOSIC implementation\cite{PRP14} employs the potential mixing as well, but it was not applied to the SIC potentials. We extend the mixing to the SIC part in this work in order to improve the robustness of the algorithm. Instead of potential mixing, we choose to apply the Broyden mixing\cite{J88} to the $\tilde{H}$ matrix directly for its simplicity. We obtain similar results and number of SCF iterations for both the potential mixing and the Hamiltonian mixing methods.

\section{Results}
All the calculations present in this paper are done with a modified version of NRLMOL\cite{PJ90,JP90,PPKP00}, which is an electronic structure code based on Gaussian-type basis sets. We use the local spin density approximation (LSDA)\cite{KS65,PW92} for exchange and correlation because of its simplicity, but there is no difficulty in using other energy functionals together with SC-FLOSIC. The forces on the FODs are converged to 0.001 a.u.

To demonstrate the effect of self-consistency, we apply SC-FLOSIC to the 11 small molecules that were calculated using non-SC-FLOSIC in Ref. \cite{PRP14} plus CH$_3$OH. The results are listed in Table \ref{table:res}. FODs in Ref. \cite{PRP14} are determined manually and not fully optimized, so some of the non-SC-FLOSIC results listed in Table \ref{table:res} are different from those in Ref. \cite{PRP14}.

\begin{table}
\begin{tabular}{lccccccc}
\hline\hline
Mol. & \multicolumn{3}{c}{Total (Hartree)} & \multicolumn{4}{c}{Atomization (eV)}\\
 & LSDA & non-SC & SC & LSDA & non-SC & SC & Exp.\\
\hline
N$_2$ & -108.692 & -109.842 & -109.856 & 11.53 & 10.22 & 10.19 & 9.76\\
O$_2$ & -149.332 & -150.736 & -150.757 & 7.48 & 5.10 & 5.14 & 5.12\\
CO & -112.471 & -113.635 & -113.648 & 12.91 & 11.18 & 11.10 & 11.11\\
CO$_2$ & -187.273 & -189.096 & -189.121 & 20.37 & 16.27 & 16.24 & 16.56\\
C$_2$H$_2$ & -76.625 & -77.594 & -77.607 & 19.91 & 18.97 & 18.92 & 16.86\\
LiF & -106.702 & -107.716 & -107.734 & 6.75 & 5.64 & 5.72 & 5.97\\
H$_2$ & -1.1251 & -1.1745 & -1.1758 & 4.57 & 4.81 & 4.79 & 4.48\\
Li$_2$ & -14.724 & -15.050 & -15.053 & 1.02 & 1.02 & 0.96 & 1.04\\
CH$_4$ & -40.109 & -40.667 & -40.676 & 19.77 & 19.64 & 19.62 & 17.02\\
NH$_3$ & -56.107 & -56.761 & -56.772 & 14.60 & 14.45 & 14.45 & 12.00\\
H$_2$O & -75.909 & -76.665 & -76.677 & 11.53 & 10.69 & 10.69 & 9.51\\
CH$_3$OH & -114.84 & -116.13 & -116.15 & 25.30 & 24.85 & 24.83 & 20.85\\
MAE & & & & 2.13 & 1.16 & 1.14\\
MARE & & & & 18\% & 8.5\% & 8.7\%\\
\hline\hline
\end{tabular}
\caption{Total energies and atomization energies of 12 molecules calculated with SC-FLOSIC and non-SC-FLOSIC. We use the experimental geometries in all calculations. The experimental atomization energies are obtained from Ref. \cite{CCCBDB}. Zero-point vibrational energies have been removed from the experimental values. The mean absolute error (MAE) and the mean absolute relative error (MARE) for the atomization energies are also listed.}
\label{table:res}
\end{table}

The self-consistency lowers the total energies systematically as expected. The effect of the self-consistency is small for the small molecules in Table \ref{table:res}. SC-FLOSIC does not further improve the atomization energies over those of non-SC-FLOSIC. To obtain more accurate atomization energies, one has to use more sophisticated energy functionals than the LSDA.

\begin{figure}
\centering
\includegraphics[width=0.75\columnwidth]{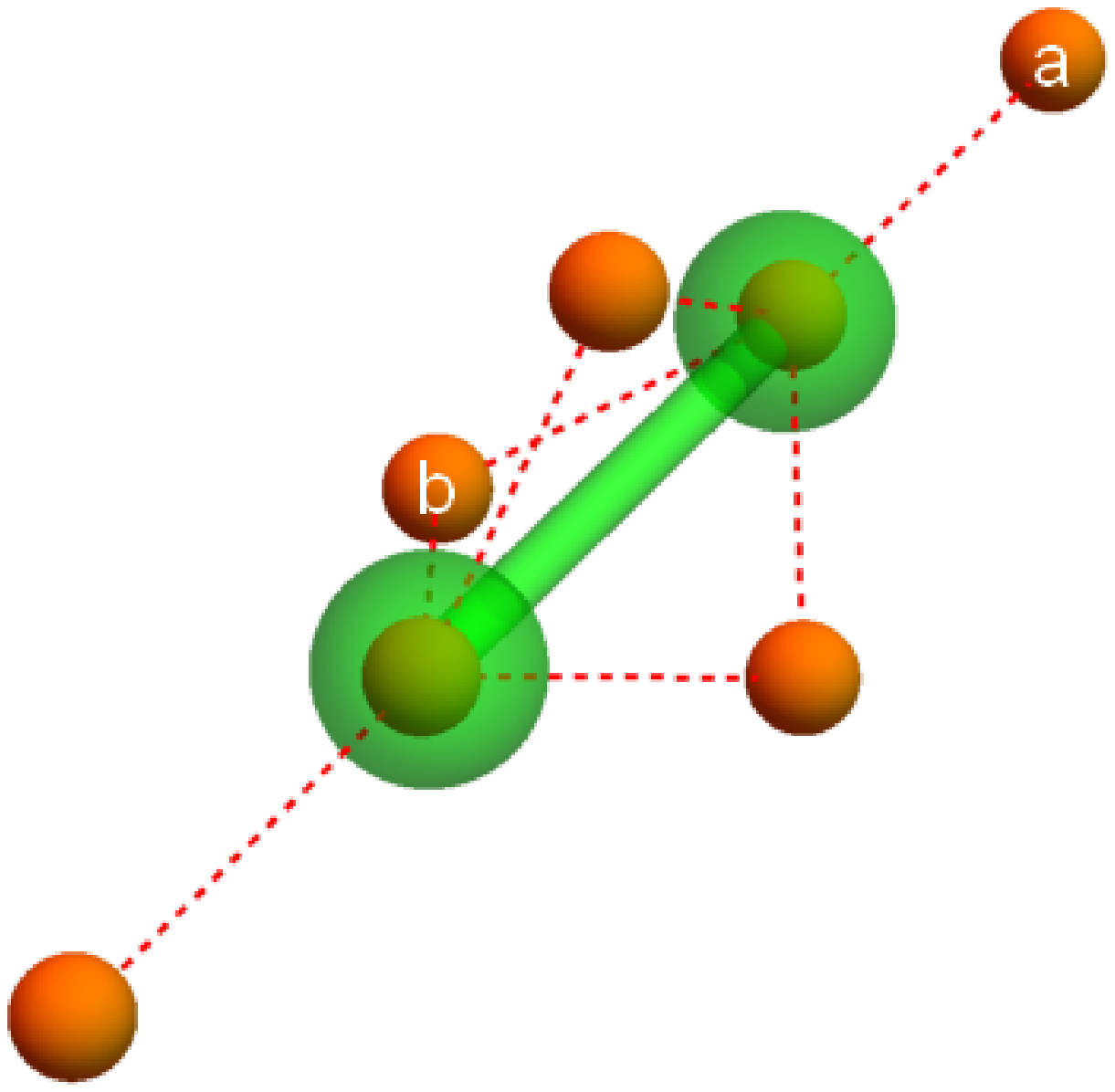}\\
(a)\\
\includegraphics[width=0.75\columnwidth]{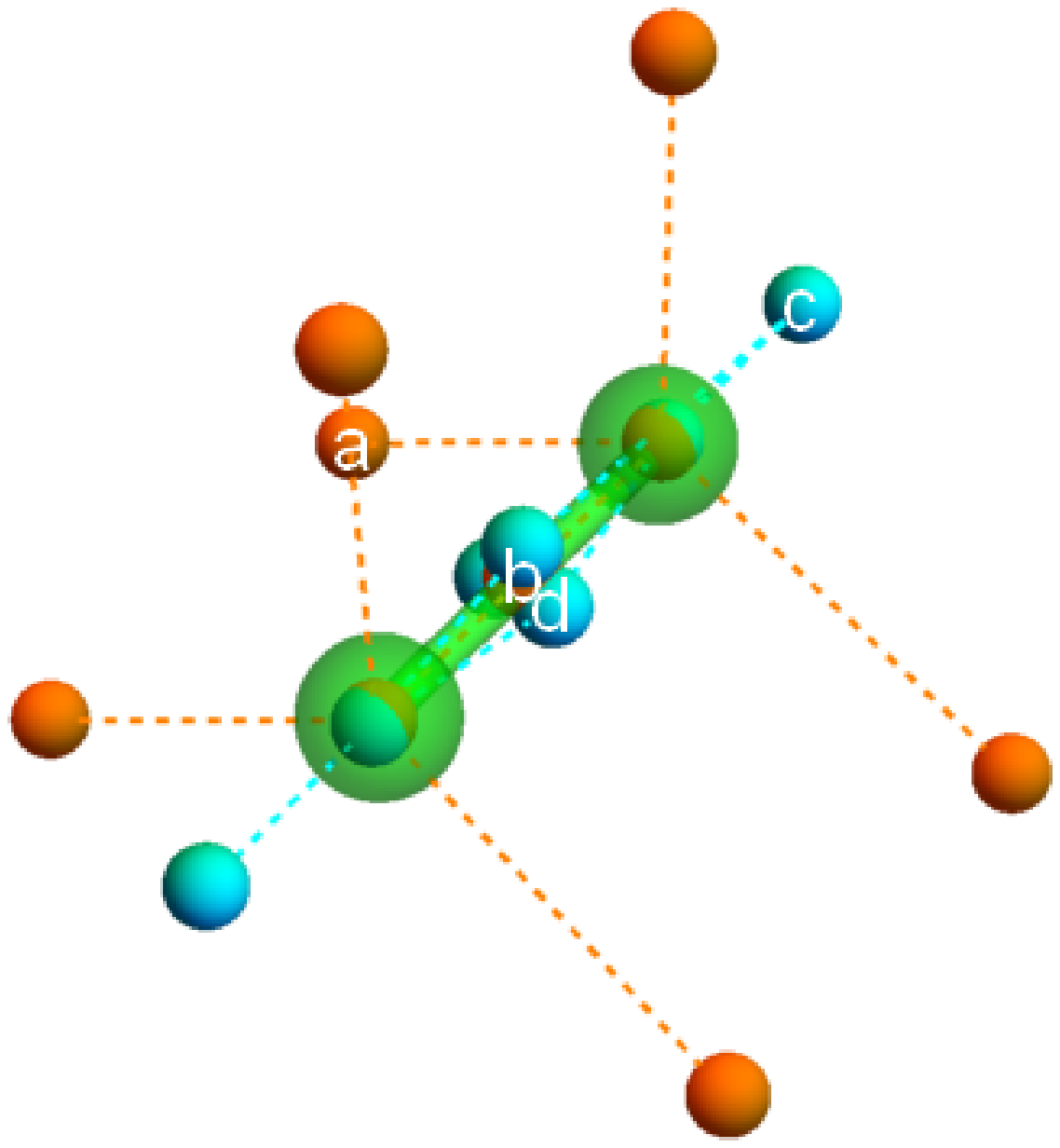}\\
(b)\\
\includegraphics[width=0.75\columnwidth]{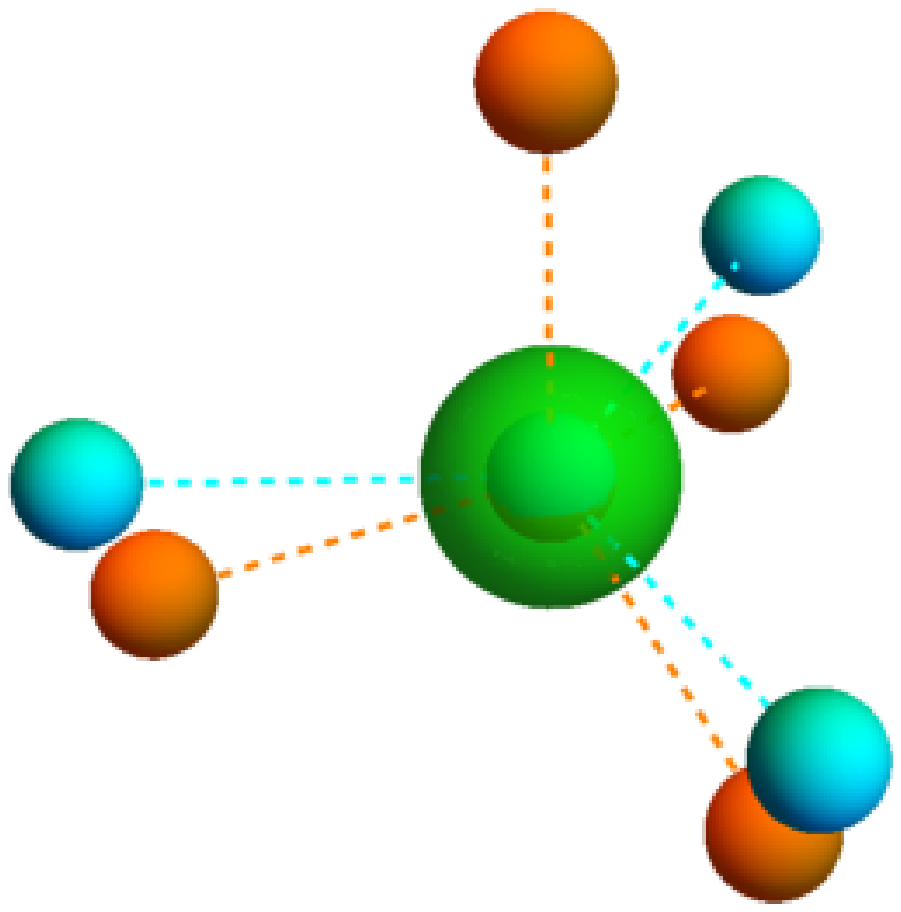}\\
(c)
\caption{(color online) FODs for (a) N$_2$, (b) O$_2$, and (c) F atom. Green transparent spheres represent atoms, brown balls represent the FODs in the spin-unpolarized case and of spin-up electrons, and cyan balls represent the FODs of spin-down electrons. Dashed lines are visual aids showing the relative positions of the FODs. The labels on the FODs are those of the FLOs plotted in Fig. \ref{fig:FO:N2} and \ref{fig:FO:O2}.}
\label{fig:FOD}
\end{figure}

\begin{figure}
\begin{minipage}[t]{0.48\columnwidth}
\centering
\includegraphics[width=\textwidth]{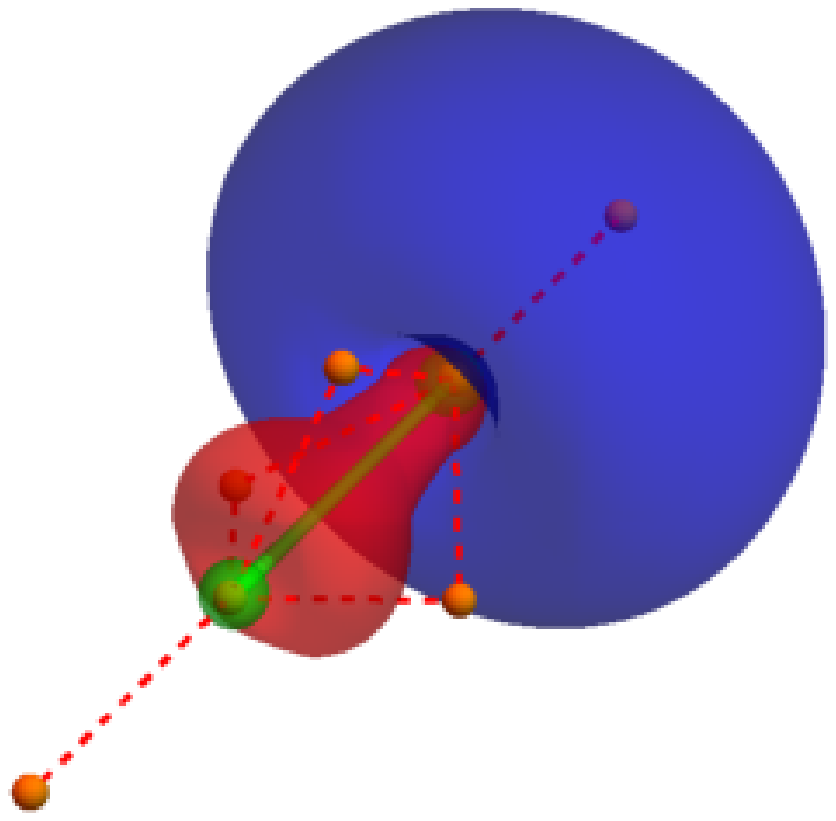}\\
(a)
\end{minipage}
\hfill
\begin{minipage}[t]{0.48\columnwidth}
\includegraphics[width=\textwidth]{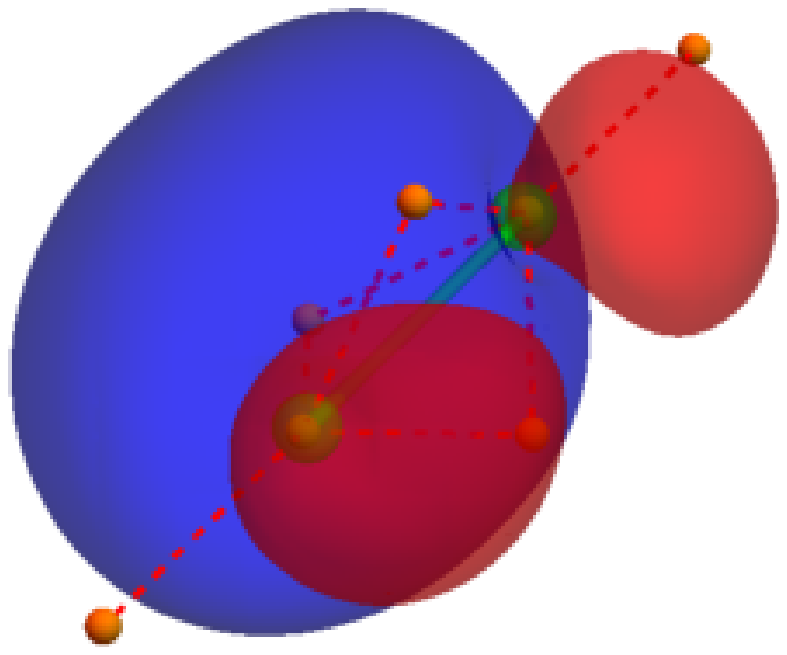}\\
(b)
\end{minipage}
\caption{(color online) The FLOs of N$_2$ corresponding to (a) the lone pairs and (b) the triple bond. Isosurfaces of $\pm0.08$ (atomic units) are plotted. The FODs of the plotted FLOs are shown in Fig. \ref{fig:FOD}(a). The FLOs corresponding to the $1s$ electrons have similar shapes as atomic 1s orbitals and are not plotted. Other FLOs are related to the plotted ones by symmetry. The atomic positions and FODs are also plotted as a visual aid.}
\label{fig:FO:N2}
\end{figure}

\begin{figure}
\mbox{
\begin{minipage}[t]{0.48\columnwidth}
\centering
\includegraphics[width=\textwidth]{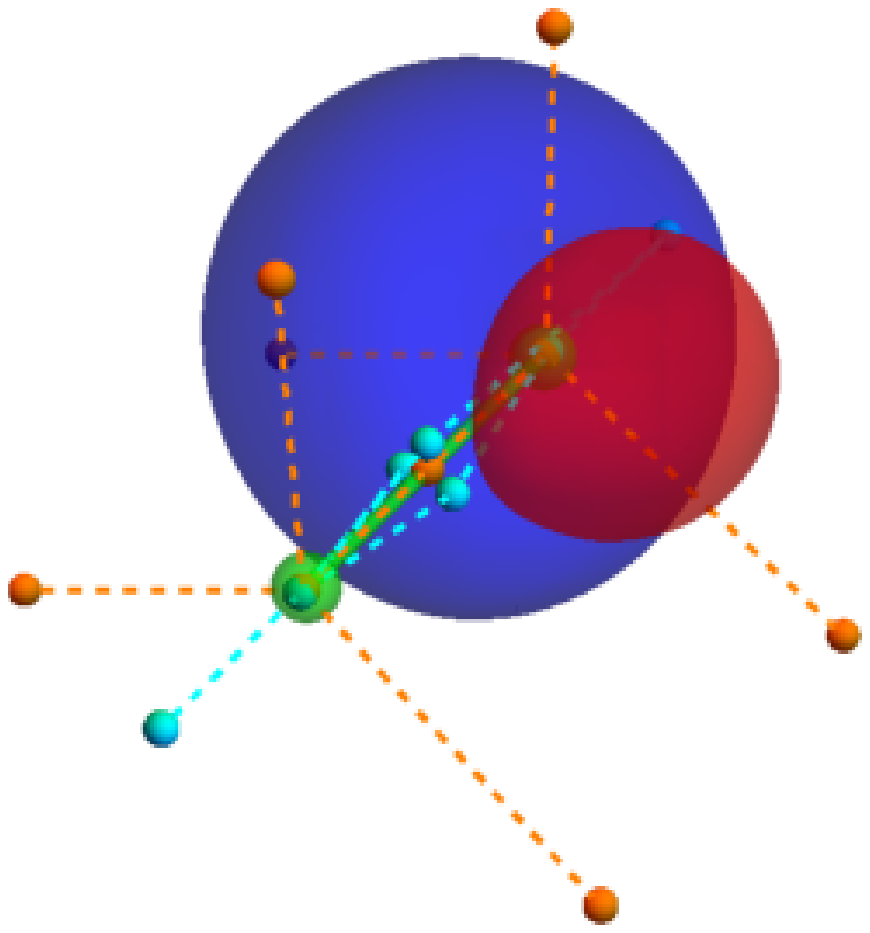}\\
(a)
\end{minipage}
\hfill
\begin{minipage}[t]{0.48\columnwidth}
\centering
\includegraphics[width=\textwidth]{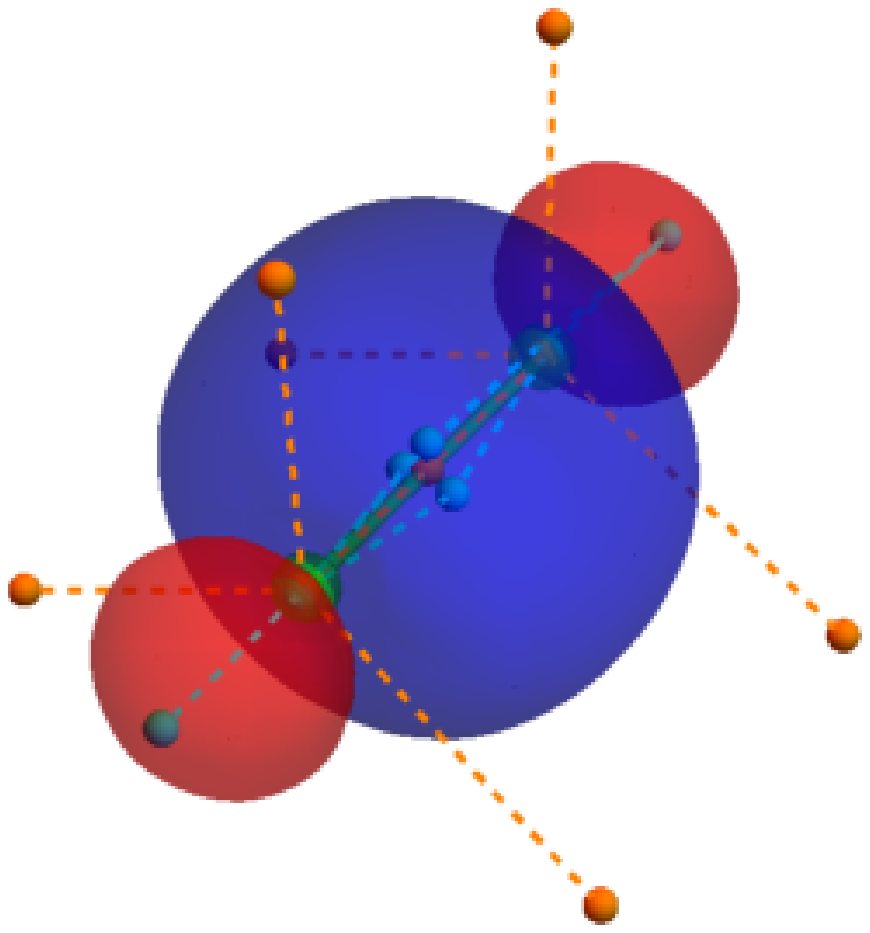}\\
(b)
\end{minipage}
}\\
\mbox{
\begin{minipage}[t]{0.48\columnwidth}
\centering
\includegraphics[width=\textwidth]{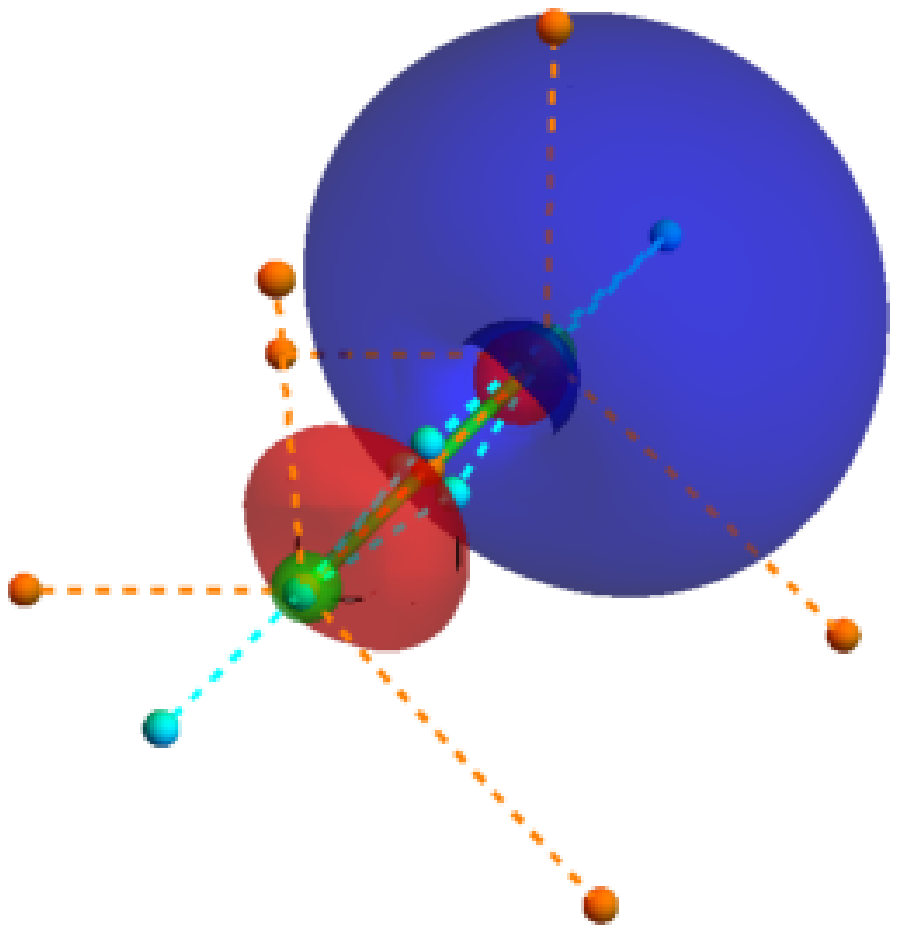}\\
(c)
\end{minipage}
\hfill
\begin{minipage}[t]{0.48\columnwidth}
\centering
\includegraphics[width=\textwidth]{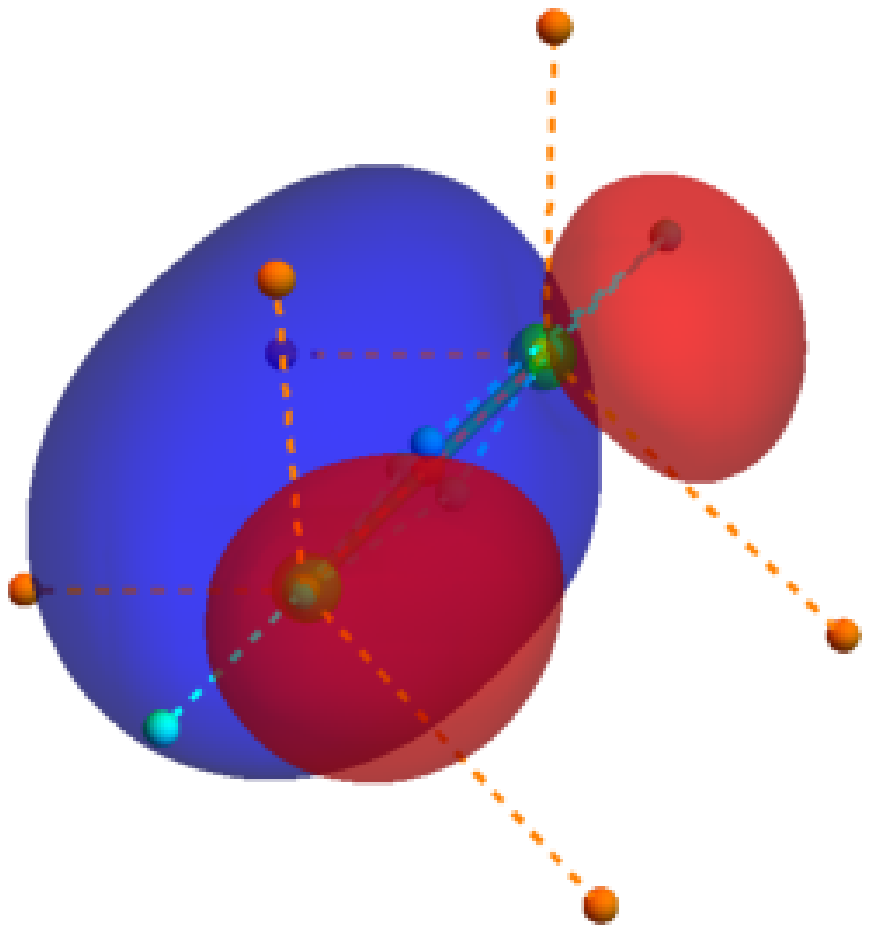}\\
(d)
\end{minipage}
}
\caption{(color online) The FLOs of O$_2$. (a)(b) are spin-up FLOs, and (c)(d) are spin-down FLOs. Isosurfaces of $\pm0.08$ (atomic units) are plotted. The FODs of the plotted FLOs are shown in Fig. \ref{fig:FOD}(b).The FLOs corresponding to the $1s$ electrons have similar shapes as atomic 1s orbitals and are not plotted. Other FLOs are related to the plotted ones by symmetry. The atomic positions and FODs are also plotted as a visual aid.}
\label{fig:FO:O2}
\end{figure}

Spin-polarized calculations are needed for O$_2$ and all the atoms in the molecules of Table \ref{table:res}. The FODs for spin-polarized calculations are more difficult to guess, and a poor initial choice usually leads to numerical instabilities. Fig. \ref{fig:FOD}(a)(b) shows the the FODs for N$_2$ and O$_2$ molecules. For the closed-shell N$_2$, the FODs and the shapes of the corresponding FLOs in Fig. \ref{fig:FO:N2} have clear chemical importances, as they correspond to the 1s electrons, the lone pairs and the triple N-N bond. As pointed out in Ref. \cite{PRP14}, the FLOs can be seen as linear combinations of molecular orbitals. For example, the banana orbitals corresponding to the triple bond are linear combinations of the $\sigma_g$, $\pi_x$ and $\pi_y$ molecular orbitals. For open-shell systems such as O$_2$, the link between FODs and the chemical bonds is more elusive. The FODs shown in Fig. \ref{fig:FOD}(b) do not appear to have clear chemical importance. Most of the spin-up and spin-down FODs are not paired as they are in the spin-unpolarized cases, and the FLOs of different spin in Fig. \ref{fig:FO:O2} have very different shapes from each other.

For atoms of the first three rows of the periodic table, the FODs are connected to the hybridization of the $s$-$p^n$ orbitals with n= 1,2, or 3 depending upon $p$ filling.\cite{PL88} As for the case of most generalized-gradient approximations, FLOSIC as well as the versions of the PZ functional which account for unitary transformations \cite{HHL83,PHL84,PHL85,PL88,FPM93,STW93,AF95,GND00,PAZ01,VS04,VS05,PASS07,KKM08,PSB08,KKJ11,HKK12,HKKK12} favors integer occupied $p$ valences over perfectly spherical atoms with fractionally occupied $p$ valences. However, especially when starting new calculations, it is necessary to account for the fact that starting points derived from spherical potentials will always provide fractionally occupied starting orbitals in open-shell systems.  To address this issue, we use starting configurations with non-spherical symmetry in calculations. As in the case of antiferromagnetic starting potentials, which are commonly used in molecular magnet calculations and molecular bond-breaking calculations, symmetric solutions would still emerge if they are indeed the lowest-energy solution. Fig. \ref{fig:FOD}(c) shows the FODs of the F atom. The three spin-up $p$ orbitals are fully occupied, and the spin-up FODs form a tetrahedron according to the $sp^3$ hybridization. Two spin-down $p$ orbitals are occupied, and the spin-down FODs form a triangle according to the $sp^2$ hybridization. The shapes of the corresponding localized orbitals also resemble $sp^3$ and $sp^2$ hybridized orbitals respectively.

For the systems in Table \ref{table:res}, the optimal FODs of the SC calculation and the non-SC calculation have the same general spatial distribution. The FODs corresponding to core electrons and chemical bonds do not change much from the SC optimal positions in the non-SC calculation, but the FODs corresponding to lone pairs for N$_2$ and CO are strongly affected by self-consistency. For the CO molecule, the distance between the non-SC FOD and the SC FOD corresponding to the lone pair on the O atom can be as long as 0.68\AA, and the corresponding non-SC and SC FLOs have visible differences.

The geometry optimization with SC FLOSIC is possible, but deriving and implementing the correct Hellmann-Feynman forces on atomic positions has not yet been completed. Nevertheless, we tested geometry optimization using LSDA forces while using the SC FLOSIC total energy to check convergence. We have limited success for very simple molecules such as H$_2$ and Li$_2$, since the LSDA forces coincides with FLOSIC forces in these simple systems. The correct FLOSIC forces are needed for more general cases. We also find that the optimization of FODs can be merged with the geometry optimization to be more efficient. The initial FODs have to be close to the correct positions in such a combined optimization, or it may not converge.

\begin{figure}
\includegraphics[width=\columnwidth]{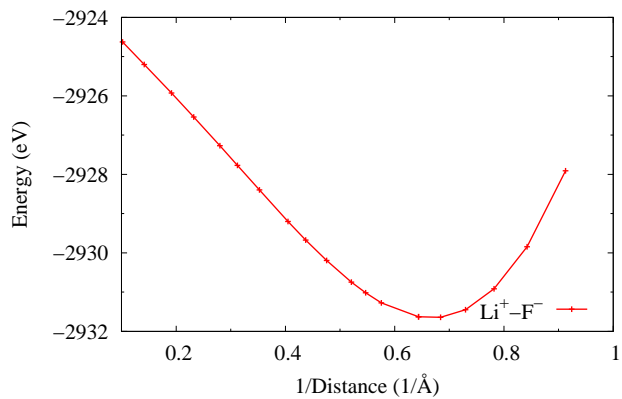}
\caption{SC FLOSIC dissociation curve of the LiF molecule for dissociation into Li$^+$ and F$^-$ ions. We use the inverse of the distance $(1/R)$ as the horizontal axis to show the $-1/R$ long range behavior of the curve.}
\label{fig:LiFionic}
\end{figure}

The SC FLOSIC dissociation curve of the LiF molecule is plotted in Fig. \ref{fig:LiFionic}. Within most approximations of density functional theory, it is generally a challenge to calculate the charge transfer energy between two fragments without introducing a constraint. For example, constrained orthogonality calculations have been introduced by Baruah and Pederson\cite{BP09} and by Ziegler and collaborators\cite{ZSKA09} which allow one to calculate an excited charge-transfer state, provided that the ground-state of the system within the approximation to DFT is qualitatively correct. However, in many cases, systems composed of two fragments are found to be electronically unstable for either two neutral separated fragments or two oppositely charged cation/anion fragment pairs\cite{S74,BS97,CCMD01,GKC04,RPCV06,CSPB10}, and rare cases where this is not a qualitative problem have been demonstrated in Ref. \cite{OBMP03}.  The LiF molecular potential energy curve provides a good example of this problem\cite{OBMP03,VSP07,MKK11}. In this case, LSDA total energies as functions of charge on the separated systems predict that the ground state of a separate Li and separated F has approximately 2.3 units of charge on the Li and 9.7 units of charge on the F (this result depends slightly on choice of functional).  Here we address the calculation of the full charge-transfer state of LiF (cation/anion pair) which can be automatically determined within FLOSIC.  In conjunction with this method and the constrained charge-transfer method of Baruah and Pederson\cite{BP09}, it then becomes possible to calculate the lower-energy neutral state as well.

It was difficult to generate the curve in Fig. \ref{fig:LiFionic} in non-SC FLOSIC, since the calculation can converge to either the neutral Li and F atoms, or to Li$^+$ and F$^-$ ions, and making sure the ground states at all separations are consistent is hard. This problem is alleviated in SC FLOSIC since we are able to solve for localized orbitals directly, so that it is easy to check the charged state of the atoms in each iteration.

\section{Conclusion}
SIC is a straightforward way to fix many of the flaws of semilocal xc energy functionals. Localized orbitals are a must to properly carry out SIC calculations. The localization equation Eq. \parref{eqn:localizationEquation} is the most general way of determining the localized orbitals, but it requires an overwhelming amount of calculation. FLOSIC simplifies the problem by reducing the number of parameters by an order of magnitude. Despite this simplification, the computational cost of FLOSIC is still high if one needs to solve Eq. \parref{eqn:localizedOrbitalSIC} directly, since the Lagrange multiplier matrix $\lambda_\sigma$ is no longer diagonal as in KS equations, and the Hamiltonian for each orbital is different. Ref. \cite{PRP14} circumvents this problem by effectively converting the problem to a regular eigenvalue problem, but the results obtained this way are not
fully self-consistent.

In this work we solve \parref{eqn:localizedOrbitalSIC} with a novel Jacobi-like numerically exact iterative algorithm, and the results don't suffer from the non-self-consistent problem associated with the method of Ref. \cite{PRP14}. The newly developed algorithm only requires $N_\sigma\times M$ extra calculations of matrix elements compared with the KS equation, instead of $N_\sigma\times M^2$ as suggested in Eq. \parref{eqn:localizedOrbitalSIC}. Compared with the $N_\sigma\times N_\sigma$ extra matrix elements used in non-SC FLOSIC, we managed to include self-consistency with marginal increase in the computation cost.

The effect of the self-consistency is small for the total energies of the simple molecules tested in this work, and the accuracy of the FLOSIC method is confirmed. The changes in the SIC total energy from the non-SC values is about $1\%\sim2\%$, and the changes in the atomization energies are smaller. The FODs mostly retain their positions as in the non-SC FLOSIC, but in some cases they show significant changes, indicating that properties depending on the density matrix may also have changed significantly. The atomization energies for both the non-SC and SC FLOSIC show vast improvement due to the removal of self-interaction, and higher accuracy can be expected from semilocal energy functionals specifically designed for SIC, or perhaps from strongly-constrained semilocal functionals like SCAN\cite{SRP15}.

This paper has developed a self-consistent approach for solving the FLOSIC equations in the representation of the FLO. We assumed the existence of canonical orbitals throughout the paper, but we must point out that, as discussed in Ref. \cite{PBKB16}, a rigorous derivation of the Kohn-Sham-like Hamiltonian for the canonical orbitals of the FLOSIC formulation is still needed. By observation, we find that Eq. (7) appears to be a good approximation to this Hamiltonian, but we also find that the magnitude of its eigenvalue for the highest occupied orbital overestimates the experimental ionization potential. This might be due to the approximation to the Hamiltonian, but it can also be due to the approximated energy functional. Further study is needed to solve this problem.

\section*{Acknowledgements}
ZY is currently supported by Science Challenge Project No. TZ2016003 (China). ZY and JP were supported in part by the National Science Foundation(Grant Nos. DMR-1305135 and DMR-1607868, with contributions from Chemical Theory, Modeling and Computation). Support from the Office of Naval Research grant N00014-16-1-2464 graciously acknowledged.


%

\end{document}